\documentclass[aps,prl,twocolumn,amsmath,showpacs,letterpaper,floatfix]{revtex4}

\newcommand{\bra}[1]{\left\langle #1\right|}
\newcommand{\ket}[1]{\left| #1\right\rangle}

\newcommand{\ieac}[0]{{\it et al., }}
\newcommand{\eeqref}[1]{Eq.~(\ref{#1})}

\usepackage{graphicx,amsmath,amssymb}

\begin{document}

\title{Quantum-optical state engineering \\up to the two-photon level}

\author{Erwan Bimbard$^{1,2}$, Nitin Jain$^{1}$, Andrew MacRae$^{1}$, A.I. Lvovsky$^{1}$}

\affiliation{$^{1}$Institute for Quantum Information Science, University of Calgary, Calgary, AB, Canada T2N 1N4}
\affiliation{$^{2}$D\'epartement de Physique, Ecole Normale Sup\'erieure, 24 rue Lhomond 75005 Paris}
\email{lvov@ucalgary.ca}



\begin{abstract}
We propose and experimentally verify a scheme to engineer arbitrary states of traveling light field up to the two-photon level. The desired state is remotely prepared in the signal channel of spontaneous parametric down-conversion by means of conditional measurements on the idler channel. The measurement consists of bringing the idler field into interference with two ancilla coherent states, followed by two single-photon detectors, which, in coincidence, herald the preparation event. By varying the amplitudes and phases of the ancillae, we can prepare any arbitrary superposition of zero- one- and two-photon states.
\end{abstract}

\maketitle

Modern quantum information science relies upon two key concepts -- superposition and entanglement. As fundamental tenets of quantum mechanics, they govern the way information can be shared, transferred or measured. 
This information lives in a Hilbert space, in the form of a \emph{quantum state}. Consequently, the ability to actively control the coherent dynamics of a quantum state is paramount for quantum information technology. This task forms the essence of quantum state engineering (QSE). In the optical domain, a widely-used approach for QSE involves \emph{generating} a ``primitive'' quantum state and then \emph{manipulating} it, for example, by bringing it into interaction with an ancillary system. Employing appropriate measurements, the ancilla is then traced out, leading to reduction of the overall system to the desired target state, ready to be \emph{detected} and \emph{characterized}.


In modern quantum optics, the ``primitive'' is commonly the state of correlated photon pairs produced in spontaneous parametric down conversion (SPDC). Conditional photon detection on one or both channels is then employed to produce the state of interest. This technique has been successfully applied to engineer complex entangled states of dual-rail optical qubits \cite{kok07}, albeit mostly in a postselected manner: we do not know that the state has been prepared until it is detected and destroyed. A variety of single-mode states have been prepared without resorting to postselection: photon number (Fock) states $\ket{1}$ \cite{spf} and $\ket{2}$ \cite{tpf06}, single-rail optical qubits \cite{resch02,catalysis}, photon-added states \cite{photonadded} and the ``Schr\"odinger kitten'' state \cite{schcat06}. However, engineering of \emph{arbitrary} quantum states of light has not yet been demonstrated.

A fundamental impediment to optical QSE arises because equidistant energy levels of a harmonic oscillator (such as an electromagnetic mode) cannot be individually accessed using classical control signals. Outside the optical domain, this challenge has been addressed in ion traps \cite{iontrap} and high-Q microwave cavities \cite{highQcav1} by coupling the oscillator energy eigenstates to a two-level atomic or spin system \cite{LawEberly}. Very recently, arbitrary superposition of Fock states were synthesized inside a superconducting cavity by means of coupling to a Josephson phase qubit \cite{highQcav2}. However, the loss of coherence in this scheme provides a fundamental limitation to the obtainable accuracy and complexity of the prepared state.

While traveling field implementations do not suffer from this kind of decoherence and also automatically satisfy DiVincenzo's criterion of ``flying qubits'' \cite{divicenzo}, they are still beset with the problem of inefficiencies and losses. There exist a number of theoretical proposals for implementing optical QSE [reviewed in detail in \cite{Dell06}], for example, using coherent displacements and photon subtraction operations \cite{dakna99a,dakna99b,fiu05}, repeated parametric down-conversion \cite{cla01}, or continuous-variable postselection \cite{lance06}.  Here we report, for the first time, a postselection-free experiment on synthesis of single-mode coherent superpositions of Fock states up to the two-photon level with stable phase relations and a high overall efficiency.

\begin{figure}[ht]
 \includegraphics[width=\columnwidth]{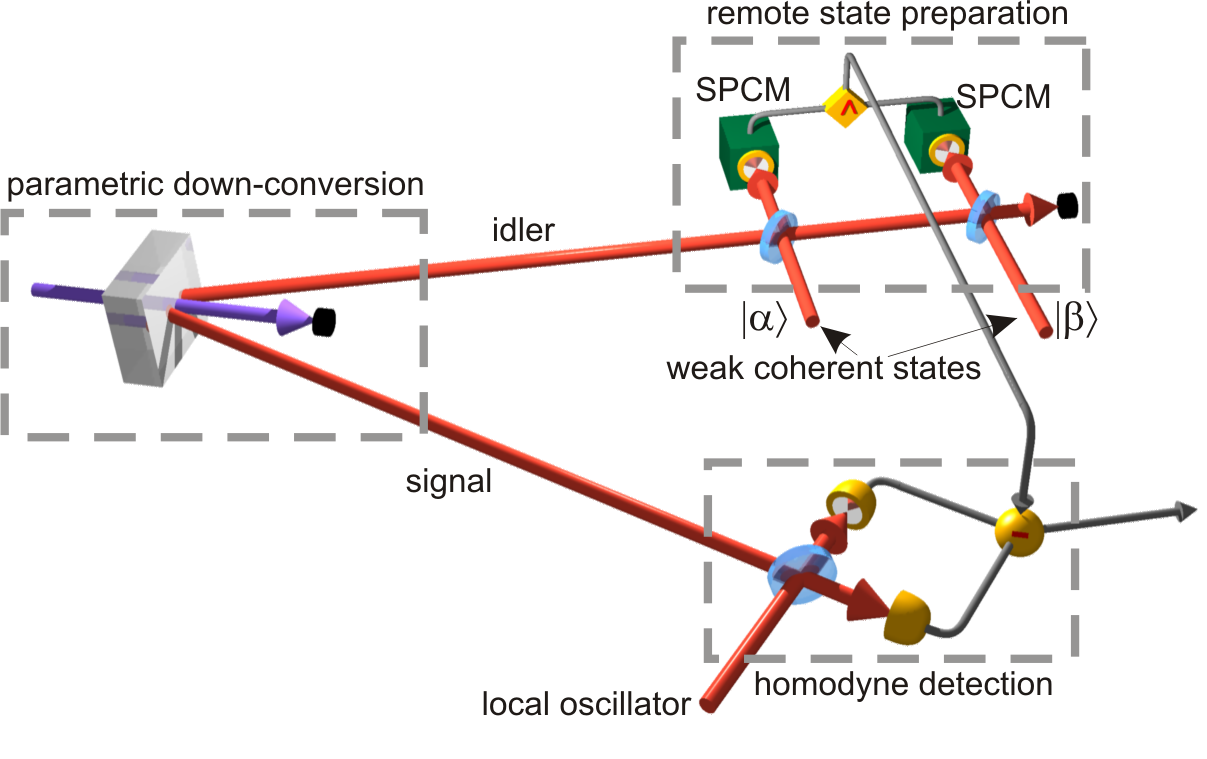}
\caption{Scheme to produce arbitrary quantum-optical states at the two photon level. The idler output of parametric down-conversion is mixed with two weak coherent states on beamsplitters, the coincidence detection from the two detectors heralds the production of the expected state in the signal channel. SPCM, single-photon counting module.}
\end{figure}

Figure 1 shows a simplified layout of our experiment. The two-mode state produced by down-conversion can be described (neglecting normalization) as
\begin{equation}\label{SPDC}
\ket\Psi=\ket{0} + \gamma \ket{1_s, 1_i} + \gamma ^2 \ket{2_s, 2_i} + O(\gamma ^3),
\end{equation}
where $\gamma ^2$ is the probability of generating a single photon pair and the signal and idler modes are represented by $s$ and $i$, respectively. The conditional measurement is performed by mixing the idler photons with two weak ancillary coherent states on symmetric beam splitters, and detecting photon arrivals in two output channels. When the two detectors ``click'' simultaneously, the non-locally prepared signal state is a superposition of $\ket{0}, \ket{1}$ and $\ket{2}$.

A qualitative explanation is as follows: each of the two trigger photons can originate either from SPDC or from the ancillary states. Since SPDC emits the same number of photons into both channels, a coincidence detection event indicates that the photon number in the signal path is 0, 1 or 2. Furthermore, if the optical modes of the ancillary and idler fields are matched spatially and temporally, the origin of the photons giving rise to registration events is fundamentally indeterminable. As a result, the prepared state $\rho$ is not simply a statistical mixture, but a \emph{coherent superposition} of the first three Fock states. The coefficients of this superposition can be controlled with the amplitude and relative phase of the two ancillae.

The commercial detectors employed in state preparation (Perkin-Elmer SPCM-AQR-14-FC) ``click'' in response to single photons, but are unable to resolve their exact number. However, the parameters $\alpha$, $\beta$, $\gamma$ are on the scale of $0.1$ \cite{olfock}, so the probability that a given ``click" has occurred in response to two or more photons entering the detector is two orders of magnitude lower than that due to a single photon. Thus it is safe to assume that all ``clicks'' are associated with single photons, and the contribution of $n\ge3$ terms in the signal channel is insignificant.

It is interesting that the entire state preparation occurs \emph{remotely}, without any manipulation of the signal channel \cite{RSP}. This is possible because the SPDC output \eqref{SPDC} is entangled. The low degree of entanglement associated with small $\gamma$ does not restrict the range of quantum states that can be engineered, but only reduces the frequency of successful preparation events.

In the limit of small $\alpha$, $\beta$ and $\gamma$, the state prepared in the signal channel can be calculated to be \begin{equation}\label{psi}
\ket\psi=a_{0} \ket{0} + a_{1} \ket{1} + a_{2} \ket{2}
\end{equation}
with
$$
a_0=-  \frac{\alpha^2}{2\sqrt{2}} + \frac{\alpha \beta}{2};\quad a_1=\frac{\beta \gamma}{2} \ket{1} ;\quad a_2=\frac{\gamma^2}{2} \ket{2}.
$$
A close examination of \eeqref{psi} reveals an interesting feature: complete elimination of the $\ket{0}$ or $\ket{1}$ components is possible simply by blocking one of the ancilla fields. If $\alpha = 0$, the first SPCM will trigger only when the first photon comes from SPDC and thus the signal channel would always contain at least one photon. If we instead block the second coherent state, a coincidence event will occur if the two triggering photons come from the first beam splitter. Due to the Hong-Ou-Mandel effect \cite{homdip}, this cannot happen if one photon originates from SPDC and one from the coherent state $\ket\alpha$. Hence a coincidence event with $\beta=0$ cannot occur if exactly one pair was generated in SPDC.


In a complete theoretical description, several parameters must be taken into account in order to reproduce and validate our experimental results. First, remote state preparation using down-conversion leads to generic inefficiencies (due to optical losses, mismatch in the modes of the signal and the local oscillator used in detection, dark counts of detectors, inefficient photodiodes and electronic noise) \cite{spf}, which can be treated as partial absorption of the signal light. Second, imperfect mode-matching of the ancilla fields with the idler photons leads to partial distinguishability and loss of coherence among Fock terms in the prepared state. Finally, phase drifts and higher photon number components must be taken into account. 

We worked with 1.7-ps laser pulses at 790 nm emitted by a mode-locked Ti:Sapphire Coherent Mira 900 laser with a repetition rate of $\sim$76 MHz. Most of the laser energy was directed towards a LBO crystal for frequency doubling. This provided us with a strong pump beam at 395 nm, which drove SPDC in a periodically poled KTiOPO$_4$ crystal. This crystal was phase-matched for type II down-conversion in which the two output modes were spatially and spectrally degenerate but had orthogonal polarizations. They were then separated spatially on a polarizing beam splitter.
Characterization of the signal field was performed using homodyne tomography \cite{htreview}, with the local oscillator (LO) obtained from the main output of the laser.

The ancilla fields also originated from the master laser. As evident from \eeqref{psi}, stability of the relative phase $\arg(\alpha)-\arg(\beta)$ of these states was critical for preserving coherence among different number states. The required stability was achieved by employing calcite beam displacers in an interferometerically stable configuration \cite{opcnotgate}. The relative phase was found to stray by less than 0.04 radians on a typical time scale of a data acquisition run (around 30 minutes).

On the other hand, the stability of the common phase $[\arg(\alpha)+\arg(\beta)]/2$ of the two ancilla states as well as the phase of the SPDC pump field $\arg(\gamma)$ was not critically important. The only effect of the drift of these phases was to modify the optical phase of the signal state $\ket\psi$. The latter was continuously monitored in the process of homodyne detection, as we describe below.

The homodyne detector employed two Hamamatsu S5972 photodiodes (85\% quantum efficiency) and featured a 90-MHz bandwidth, which permitted time-resolved quadrature measurements at the LO pulse repetition rate \cite{olfock}. Time-integrated photocurrent samples associated with each LO pulse were normalized with respect to the vacuum state, yielding quadrature data of the state measured. For each state reconstruction, 50000 samples were typically acquired.

In order to reconstruct the state, we also needed to know the phase of the signal in reference to LO for each quadrature sample. This information could be deduced by analyzing time-dependent quadrature statistics, because at any given phase $\phi$ the average quadrature value behaves as $\langle Q_\phi \rangle \propto \sin \phi$. However, in our experiment the phase varied on a time scale of $\sim 0.5$ rad/s, so the coincidence-triggered data, acquired at a rate of 20--200 Hz, were insufficient for reliable phase reconstruction. Therefore we additionally acquired the homodyne signal triggered by events from both individual single-photon detectors, which occurred at a much higher rate of 25--100 kHz. The signal states prepared by these events are phase-sensitive coherent superpositions of Fock states $\ket 0$ and $\ket 1$. Average quadrature values of these states over $\sim 0.06$ s periods were used to determine the LO phase at each moment in time. The phase-quadrature pairs from the acquisition were then fed to a maximum-likelihood reconstruction algorithm \cite{MaxLik1} to estimate the density matrix of the signal state.

\begin{figure}[ht]
\begin{center}
 \includegraphics[width=\columnwidth]{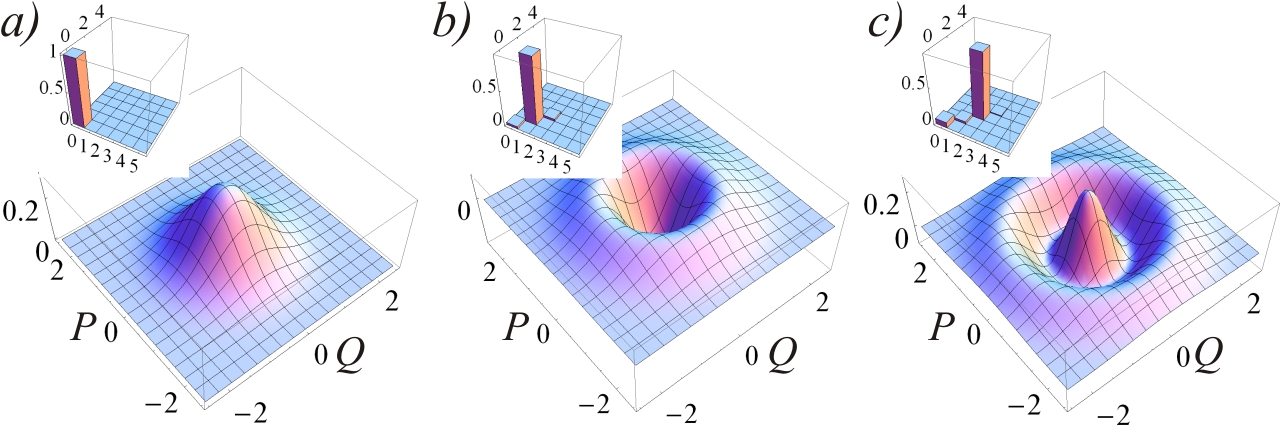}
\caption{Results for Fock states $\ket{0}$ (a), $\ket{1}$ (b), and $\ket{2}$ (c). The Wigner function and the density matrix (absolute values) are displayed. All state reconstructions feature correction for 55\% detection efficiency.}
\end{center}
\end{figure}


Each acquisition run began with preparation and reconstruction of single-photon Fock states, by blocking the ancilla fields (i.e. $\alpha = \beta = 0$) and triggering the quadrature acquisition by individual SPCMs. This allowed us to test the performance of our setup and determine its overall efficiency. The latter was typically found to be between 0.52 and 0.55, in agreement with the expected value of 0.54 calculated as a cumulative quantum efficiency of all parts of the experiment \cite{olfock}. In subsequent applications of the state reconstruction algorithm, we corrected for this known inefficiency using a value of $\eta=0.55$  \cite{MaxLik1}. In order to further test our procedure, we have prepared and reconstructed the three basis Fock states (Fig.~2).

\begin{figure}[ht]
\begin{center}
 \includegraphics[width=\columnwidth]{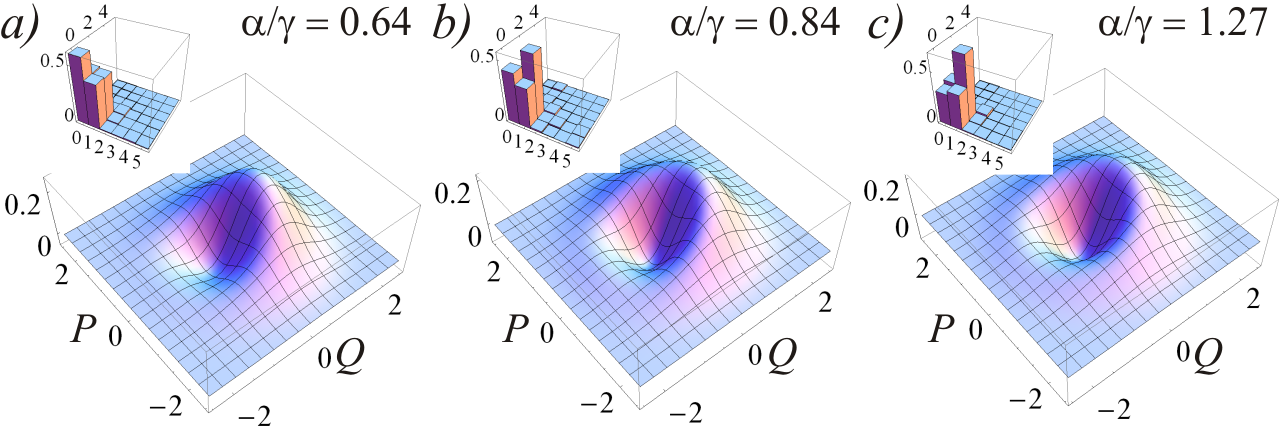}
\caption{Reconstructed superpositions of states $\ket 0$ and $\ket 1$. The single-photon fraction increases from left to right.}
\end{center}
\end{figure}

\begin{figure}[ht]
\begin{center}
 \includegraphics[width=\columnwidth]{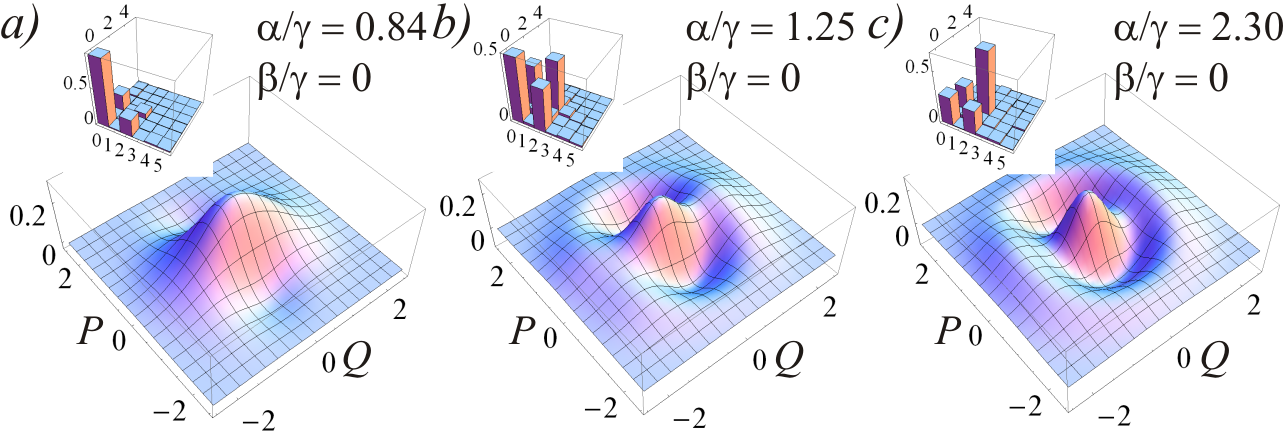}
\caption{Reconstructed superpositions of states $\ket 0$ and $\ket 2$. The two-photon fraction increases from left to right.}
\end{center}
\end{figure}

\begin{figure}[ht]
\begin{center}
 \includegraphics[width=\columnwidth]{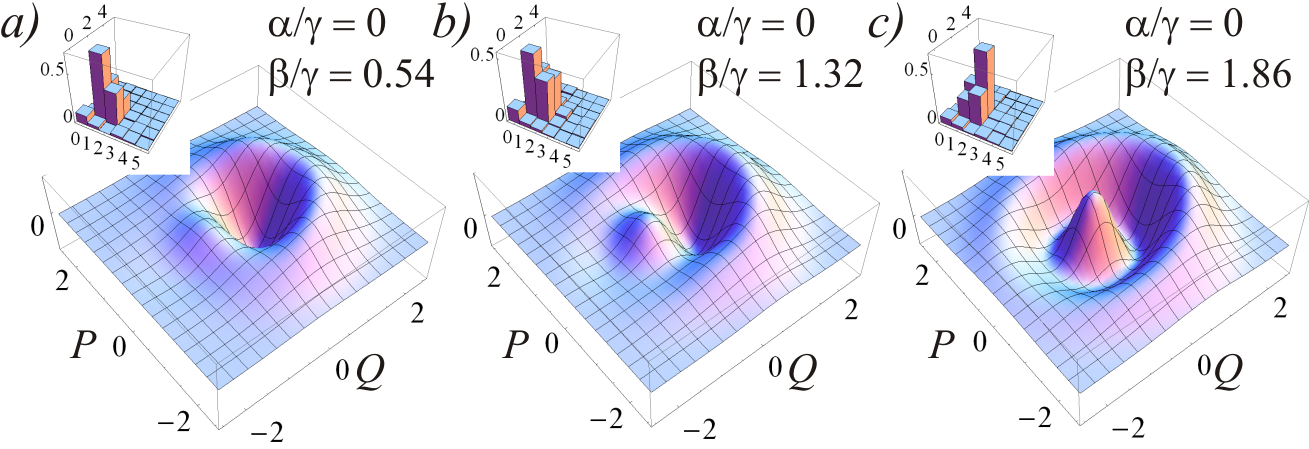}
\caption{Reconstructed superpositions of states $\ket 1$ and $\ket 2$. The two-photon fraction increases from left to right.}
\end{center}
\end{figure}

Figures 3, 4, and 5 display coherent superpositions of states $\ket 0$ and $\ket 1$, $\ket 0$ and $\ket 2$, $\ket 1$ and $\ket 2$, respectively. The states in Fig.~3 were prepared by conditioning the quadrature acquisition on click events from only the first SPCM in the presence of a single ancilla field $\ket\alpha$. With this setting, the heralded signal state was $\ket\psi=\alpha\ket 0+\gamma\ket 1$ \cite{resch02}. The states in Figs.~4 and 5, on the other hand, were prepared by blocking one of the weak ancilla inputs and triggering the homodyne acquisition on coincidence counts. The three examples in each figure contain varying weights of superposition terms, implementing gradual transition from one Fock state to another. It is interesting to note that the state in Fig.~4(a), i.e.~the vacuum state with a small contribution of $\ket 2$, is a good (95\% fidelity) approximation of the even ``Schr\"odinger kitten'' state $\ket\alpha+\ket{-\alpha}$ with $\alpha=0.60$.


Superpositions of all three Fock states are displayed in Fig.~6. In this acquisition run, the phases of $a_0$, $a_1$, and $a_2$ were set equal by adjusting the classical interference of the two ancilla fields in the non-blocked output of the second beam splitter to a constructive fringe. In the three states displayed, the values of $\beta$ and $\gamma$ were maintained approximately the same but the amplitude $\alpha$ of the first coherent state was varied. As a result, we observe a gradual transition between the vacuum state and an equal-weight coherent superposition of $\ket 1$ and $\ket 2$.

\begin{figure}[ht]
\begin{center}
 \includegraphics[width=\columnwidth]{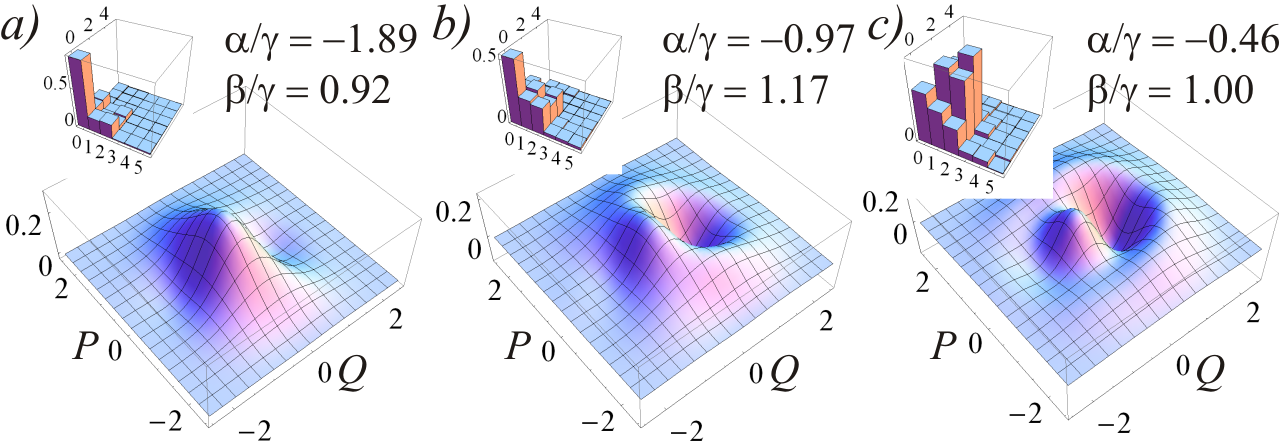}
\caption{Reconstructed equal-phase superpositions of states $\ket 0$, $\ket 1$ and $\ket 2$. In all plots, $a_1\approx a_2$; the vacuum amplitude $a_0$ decreases from left to right.}
\end{center}
\end{figure}

\begin{figure}[ht]
\begin{center}
 \includegraphics[width=\columnwidth]{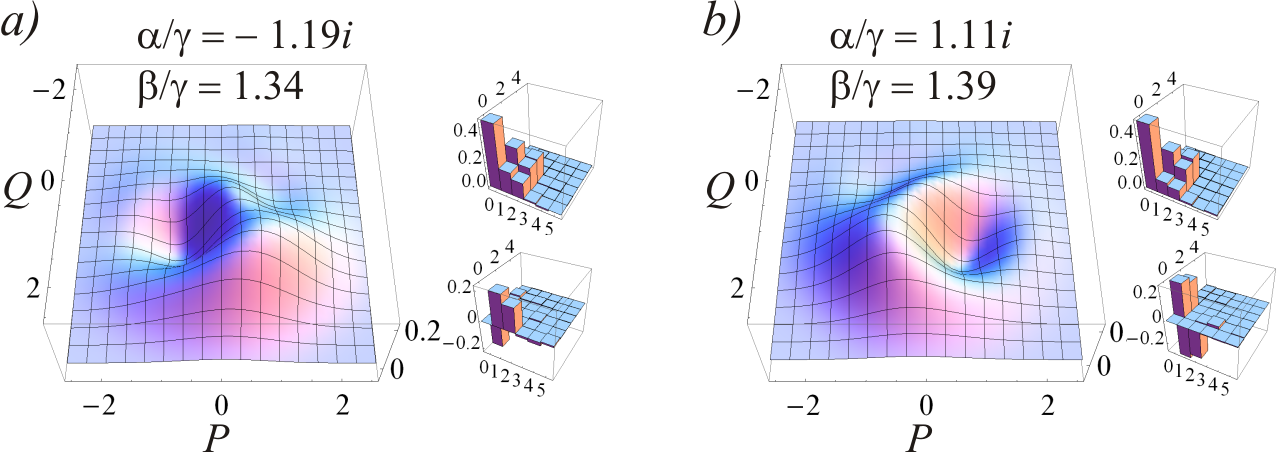}
\caption{Superpositions of states $\ket 0$, $\ket 1$ and $\ket 2$. The magnitudes of $a_0$,$a_1$, and $a_2$ in (a) are approximately equal to those in (b). Amplitudes $a_0$ and $a_1$ are real whereas the $a_2$ are complex conjugate of each other in (a) and (b), so the Wigner functions appear as mirror reflections of each other. The real and imaginary components of the density matrix are shown for both states.}
\end{center}
\end{figure}

In all of the above examples, the zero phase reference of the reconstructed state could be chosen to make all density matrix elements real within the experimental error, leading to reflection symmetry of the reconstructed Wigner functions. This is no longer the case in our two final acquisitions, aimed to demonstrate our control over the phases of $a_i$ (Fig.~7). Here the ancillae's phases were adjusted to observe interference midway between a bright and dark fringe, on opposite fringe sides in Figs.~7(a) and (b). This implies a phase difference of $\pm\pi/2$ between $\alpha$ and $\beta$. If the optical phase of the reconstructed state is chosen so that $a_1$ and $a_2$ are real, the amplitude of $a_0$ in Fig.~7(a) is complex conjugate of that in Fig.~7(b). As a result, the two Wigner functions themselves are not symmetric, but exhibit mirror symmetry with respect to each other.


In order to estimate the quality of our state preparation, we fitted each of the reconstructed states $\hat\rho$ with \eeqref{psi}, aiming to maximize the fidelity $F=\bra\psi\hat\rho\ket\psi$. The fit parameters were restricted to agree with the experimental conditions under which each data set was taken.
The obtained best fit parameters are shown in Figs.~3-7 next to each plot. All states studied in this work exhibited at least a 76\% fidelity with the fits.


To summarize, we have experimentally demonstrated production and characterization of superpositions of Fock states $\ket{0}$, $\ket{1}$ and $\ket{2}$ using parametric down-conversion and non-local state preparation. We used two weak coherent states as ancillae, which we mixed with the SPDC idler channel in a phase-stable setting to devise a measurement that remotely prepares the desired signal state. Blocking one of the coherent states allowed us to remove the $\ket{0}$ or $\ket{1}$ component from the superposition. Extension of this method to a higher photon number domain is possible, however requires additional theoretical work as well as enhancement of experimentally accessible photon-pair sources.

This work was supported by NSERC, iCORE, CFI, AIF, Quantum\emph{Works}, and CIFAR. We thank S. Huisman and M. Lobino for assistance.


\end{document}